\documentclass[a4paper,australian,aps,prl,reprint,twocolumn,twoside,superscriptaddress,preprintnumbers,showpacs,showkeys,lengthcheck,final]{revtex4-1}

\usepackage{babel}
\usepackage{amsmath}
\usepackage{braket}
\usepackage{graphicx}
\usepackage{mathptmx}
\usepackage{bm}
\usepackage{float}

\newcommand{\tr}{\mathop{\textrm{tr}}}
\newcommand{\ee}{\mathrm{e}}
\newcommand{\ii}{\mathrm{i}}

\newcommand{\transpose}{\mathrm{T}}

\newcommand{\pisigma}{\pi\Sigma}
\newcommand{\KbarN}{\smash{\overline{\textrm{K}}}\textrm{N}}

\begin{document}

\preprint{ADP-11-28/T750}

\title{Isolating the $\bm\Lambda$(1405) in Lattice QCD}
\author{Benjamin J.\ Menadue}
\affiliation{Special Research Centre for the Subatomic Structure of Matter, School of Chemistry \& Physics, University of Adelaide, South Australia 5005, Australia}
\author{Waseem Kamleh}
\affiliation{Special Research Centre for the Subatomic Structure of Matter, School of Chemistry \& Physics, University of Adelaide, South Australia 5005, Australia}
\author{Derek B.\ Leinweber}
\affiliation{Special Research Centre for the Subatomic Structure of Matter, School of Chemistry \& Physics, University of Adelaide, South Australia 5005, Australia}
\author{M.\ Selim Mahbub}
\affiliation{Special Research Centre for the Subatomic Structure of Matter, School of Chemistry \& Physics, University of Adelaide, South Australia 5005, Australia}



\date{29 September 2011}

\begin{abstract}
The negative-parity ground state of the $\Lambda$ baryon lies surprisingly low in mass. At 1405.1 MeV, it lies lower than the negative-parity ground state nucleon, even though it has a valence strange quark. Using the PACS-CS $(2+1)$-flavour full-QCD ensembles available through the ILDG, we employ a variational analysis using source and sink smearing to isolate this elusive state. We find three low-lying odd-parity states, and for the first time reproduce the correct level ordering with respect to the nearby scattering thresholds.
\end{abstract}

\maketitle

The $J^P = {1/2}^-$ ground-state resonance of the $\Lambda$ baryon, $\Lambda(1405)$, lies anomalously low in mass. At $1405.1^{+1.3}_{-1.0}$\,MeV \cite{Nakamura:2010zzi}, it not only lies lower than the first positive-parity excited state, but also lower than the negative-parity ground state of the nucleon -- even though it has a strange valence quark. Lattice QCD is the only first-principles, non-perturbative method for investigating low-energy QCD. However, no Lattice QCD study to date has successfully identified the mass suppression associated with the $\Lambda(1405)$ \cite{Nemoto:2003ft,Burch:2006cc,Takahashi:2009bu}. For example, recent work by Takahashi and Oka \cite{Takahashi:2009bu} identified two nearly degenerate states at around 1.6\,GeV.

The CSSM Lattice Collaboration has recently developed an effective technique for investigating the low-lying nucleon spectrum, using local operators that have been smeared with gauge-invariant Gaussian smearing at both the source and sink, together with a variational analysis \cite{Mahbub:2009nr,Mahbub:2010jz}. We use the same techniques to investigate the negative-parity, spin-$1/2$ spectrum of the $\Lambda$, and report the identification of low-lying states associated with the physical states of Nature. For the first time, we identify states low enough to be associated with the $\Lambda$(1405).

The variational analysis takes advantage of the extra information found by calculating cross-correlation functions for different operators at the source and sink to isolate individual states \cite{Luscher:1990ck,Michael:1985ne}. Such an analysis is necessary as the lowest three $J^P = {1/2}^-$ states of the $\Lambda$ all lie within a 400\,MeV range, at 1405.1\,MeV, 1670\,MeV, and 1800\,MeV \cite{Nakamura:2010zzi}. Because $\mathop{\textrm{SU}}(3)$ flavour symmetry is broken by the heavier strange quark mass, all three of these states will survive until the signal is buried in noise. Hence the long-time approximation usually used to extract ground-states will only be able to resolve a mixture of these low-lying states \cite{Menadue:2011cw}.

\begin{figure}
\includegraphics{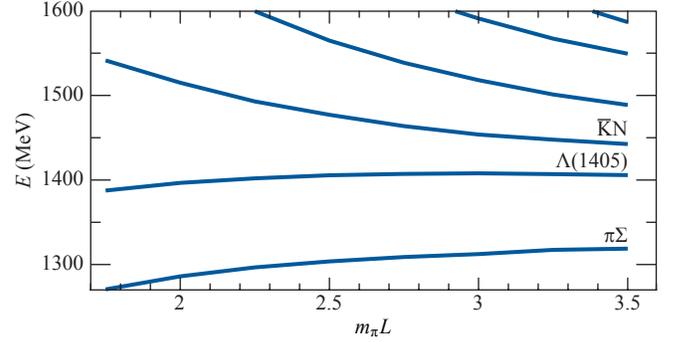}
\caption{\label{levels}(colour online) Energy spectrum of the $I = 0$, $S = -1$, $J^P = {1/2}^-$ meson-baryon sector obtained from the $\KbarN$ J\"ulich model of hadron exchange \cite{Doring:2011ip}, showing small finite-size effects for the $\Lambda(1405)$. Adapted from \cite{Doring:2011ip}.\vspace{-15pt}}
\end{figure}

In comparison to the Roper resonance, where significant finite-size effects develop through avoided level crossings between the baryon and the multi-particle scattering states, the odd-parity $\Lambda$(1405) is relatively independent of the box size. The lowest multi-particle scattering states, $\pisigma$ and $\KbarN$, lie, respectively, below and above the $\Lambda(1405)$, and do not cross for the lattice volumes depicted in Figure~\ref{levels}. As such, finite-size effects are benign in this analysis.

Consider a set of $N$ operators $\chi_i(\mathbf{x}, t)$. Using these, we can construct the $N\times N$ correlation matrix of cross-correlation functions
\begin{equation}
G^\pm_{ij}(\mathbf{p}, t) = \sum_\mathbf{x} \ee^{-\ii \mathbf{p} \cdot \mathbf{x}} \tr ( \Gamma_\pm \braket{\Omega | \chi_i(\mathbf{x}, t) \smash{\overline\chi}_j(\mathbf{0}, 0) | \Omega}), \\
\end{equation}
where the trace is taken over the (implicit) spinor indices and $\Gamma_\pm$ are the parity projection operators used to project into definite positive- or negative-parity. Working at zero-momentum, $\Gamma_\pm := (\gamma_0 \pm 1)/2$, and
\begin{align}
G^\pm_{ij}(\mathbf{0}, t) &= \sum_\mathbf{x} \tr ( \Gamma_\pm \braket{\Omega | \chi_i(\mathbf{x}, t) \smash{\overline\chi}_j(\mathbf{0}, 0) | \Omega}) \nonumber \\
&= \sum_\alpha \lambda_i^{\alpha\pm}\smash{\overline\lambda}_j^{\alpha\pm} \ee^{-m_{\alpha}^\pm t}.
\end{align}
Here, $\alpha$ enumerates the energy eigenstates of parity $\pm$ with mass $m_\alpha^\pm$, and $\lambda_i^{\alpha\pm}$ and $\smash{\overline\lambda}_j^{\alpha\pm}$ are the couplings of the operators $\chi_i$ and $\smash{\overline\chi}_j$ to these eigenstates.

With the $t$ dependence only in the exponential terms, we look for a linear combination of operators $\phi^\alpha = \sum_i v_i^\alpha \chi_i$ and $\smash{\overline\phi}^\alpha = \sum_i u_i^\alpha \smash{\overline\chi_i}$ such that
\begin{align}
G_{ij}^\pm(t_0 + \Delta t) u_j^\alpha &= \ee^{-m_\alpha^\pm \Delta t} G_{ij}(t_0) u_j^\alpha\textrm{ and}\nonumber \\
v_i^\alpha G_{ij}^\pm(t_0 + \Delta t) &= \ee^{-m_\alpha^\pm \Delta t} v_i^\alpha G_{ij}(t_0)
\end{align}
for sufficiently large $t_0$ and $\Delta t$. Rearranging this, we construct left- and right-eigensystems for the matrix $G^\pm(t_0 + \Delta t) G^\pm(t_0)^{-1}$:
\begin{align}
G^\pm(t_0)^{-1} G^\pm(t_0 + \Delta t) \mathbf{u}^\alpha &= \ee^{-m_\alpha^\pm \Delta t} \mathbf{u}^\alpha,\nonumber \\
\mathbf{v}^{\alpha\transpose} G^\pm(t_0 + \Delta t) G^\pm(t_0)^{-1} &= \ee^{-m_\alpha^\pm \Delta t} \mathbf{v}^{\alpha\transpose}.
\end{align}
Moreover, the eigenvectors $\mathbf{u}^\alpha$ and $\mathbf{v}^\alpha$ diagonalise the correlation matrix at times $t_0$ and $t_0 + \Delta t$, and allow us to construct the eigenstate projected correlation functions
\begin{equation}
G^\pm_\alpha(t) := v_i^\alpha G_{ij}^\pm(t) u_j^\alpha.
\end{equation}
We then analyse these parity and eigenstate projected correlation functions using standard effective mass techniques,
\begin{equation}
m_\alpha^\pm = \ln \frac{G^\pm_\alpha(t)}{G^\pm_\alpha(t + 1)}.
\end{equation}
More details can be found in \cite{Mahbub:2009nr,Blossier:2009kd}.

Given that the $\Lambda$ baryon lies in the centre of approximate $\mathop{\textrm{SU}}(3)$-flavour, there are a variety of operators that will couple to it, corresponding not only to the usual various Dirac structures, but also to the different possible flavour-symmetry structures. To initially investigate the spectrum, we use the so-called ``common'' operators $\chi_1^\textrm{c}(x)$ and $\chi_2^\textrm{c}(x)$ \cite{Leinweber:1990dv}, which are constructed using the terms common to both the octet and singlet operators. $\chi_1^\textrm{c}(x)$ has a Dirac structure of $(q^\transpose C\gamma_5 q)q$, while $\chi_2^\textrm{c}(x)$ has $(q^\transpose C q)\gamma_5 q$. These isospin-0 interpolators make no assumptions about the flavour symmetry, and couple to all states of the $\Lambda$ baryon. We will also consider the analogous flavour-octet and -singlet interpolators \cite{Leinweber:1990dv}. Similar to the investigation of the nucleon spectrum, we employ gauge-invariant Gaussian smearing \cite{Gusken:1989qx} at both the source and sink to create our operator basis. We consider 16, 35, 100, and 200 sweeps of smearing.

\begin{figure}
\includegraphics{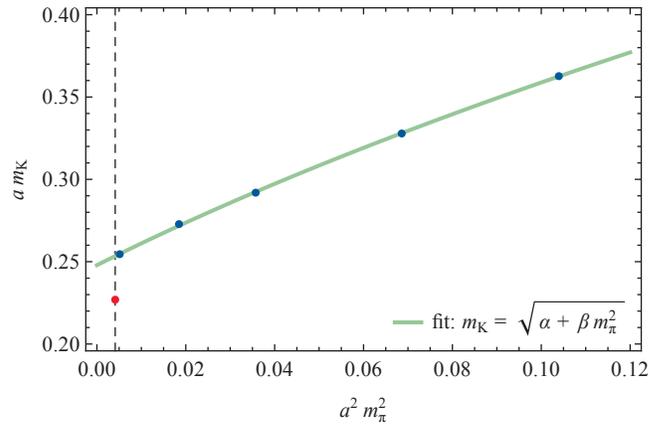}
\caption{\label{fig:kaon}(colour online) The kaon mass plotted against $m_\pi^2$ in lattice units on the PACS-CS ensembles, as quoted in \cite{Aoki:2008sm}. Extrapolating to the physical limit using $m_\textrm{K}^2 = \alpha + \beta m_\pi^2$, the kaon is approximately 60\,MeV higher than the physical kaon.\vspace{-15pt}}
\end{figure}

We use the PACS-CS $(2+1)$-flavour full-QCD ensembles \cite{Aoki:2008sm}, available through the ILDG \cite{Beckett:2009cb}. They are $32^3\times 64$ lattices with $\beta = 1.90$, giving a lattice spacing of $0.0907(13)$\,fm. There are 5 light quark masses available, with hopping parameters $\kappa_\textrm{u,d} = 0.13700, 0.13727, 0.13754, 0.13770,\textrm{ and }0.13781$, corresponding to pion masses ranging from 702\,MeV down to 156\,MeV. The strange quark mass is the same for all light quark masses, with hopping parameter $\kappa_\textrm{s} = 0.13640$. However, this is slightly too high to reproduce the physical kaon mass. Plotting the kaon mass data provided in \cite{Aoki:2008sm} against $m_\pi^2$ (Fig.~\ref{fig:kaon}) and extrapolating to the physical limit, we see that the kaon lies approximately 60\,MeV too high. We calculate the $\kappa_\textrm{s}$ required to reproduce a physical-mass kaon using two methods: first, by requiring the correct mass for the $\smash{\overline{s}}s$ pseudoscalar meson, and second, by requiring the correct kaon mass at the lightest available light quark mass through the Gell-Mann--Oakes--Renner relation. Both methods produce almost identical hopping parameters, and so we average to obtain $\kappa_\textrm{s} = 0.13694$. We first perform our variational analysis using the dynamical $\kappa_\textrm{s}$ and then again after partially quenching the strange-quark sector by shifting the strange valence $\kappa_\textrm{s} = 0.13640$ to $\kappa_\textrm{s} = 0.13694$.

\begin{figure}
\includegraphics{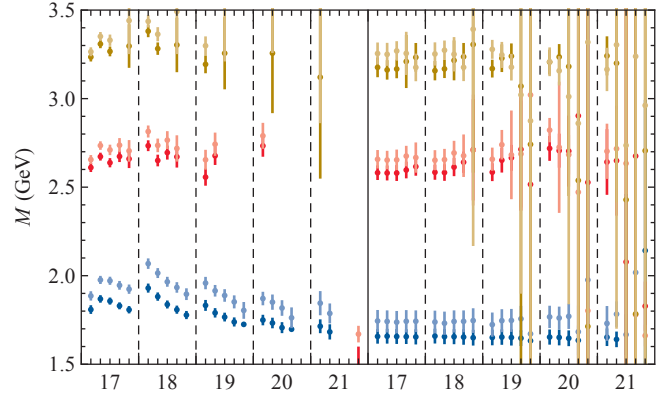}
\caption{\label{parameters}(colour online) Comparison of the mass as extracted from the eigenvalues (left) and from fitting the projected effective mass (right) with $\kappa_\textrm{u,d} = 0.13770$ for $t_0 \in \{17,\ldots,21\}$ and $\Delta t \in \{1,\ldots,5\}$. Numbers of the abscissa indicate $t_0$ with $\Delta t$ increasing within each $t_0$. Complex eigenvalues are not displayed.\vspace{-15pt}}
\end{figure}

We begin our analysis of the negative-parity $\Lambda$ by investigating the dependence of the spectrum on variational parameters $t_0$ and $\Delta t$, and operators included in the correlation matrix. The masses extracted from both the eigenvalues and from fitting the projected effective masses are examined. We consider the ensemble with $\kappa_\textrm{u,d} = 0.13727$, and select the dimension-6 operator basis comprised of $\chi_1^\textrm{c}$ and $\chi_2^\textrm{c}$ together with 16, 100, and 200 sweeps of smearing to determine the optimal variational parameters (Fig.~\ref{parameters}). While the eigenvalues show a significant dependence on the variational parameters, the fitted projected effective masses are stable and show little parameter dependence for sufficiently large $t_0$ and $\Delta t$. We select $t_0 = 18$ and $\Delta t = 2$ (with the source located at $t = 16$) as representative.

\begin{figure}
\includegraphics{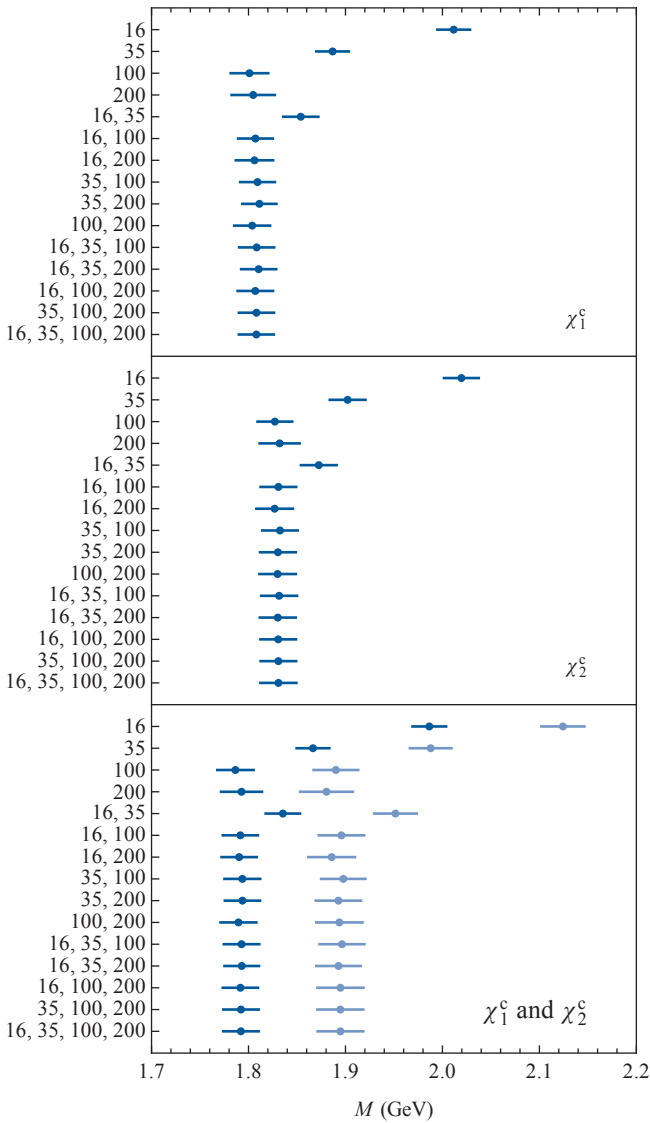}
\caption{\label{basis}(colour online) Comparison of the lowest-lying masses extracted from fitting the projected effective mass for $\kappa_\textrm{u,d} = 0.13727$ over bases formed from combinations of either or both ``common'' operators $\chi_1^\textrm{c}$ and $\chi_2^\textrm{c}$ and all smearing levels.\vspace{-15pt}
}
\end{figure}

Using these parameters, we investigate the dependence on the operator basis in Figure~\ref{basis}. We find that we require both $\chi_1^\textrm{c}$ and $\chi_2^\textrm{c}$ to resolve the lowest states; otherwise only a mixed state is extracted. We cannot resolve the three low-lying states that are present in the physical spectrum with any combination of these operators. For combinations which contain sufficiently large amounts of smearing, there is little to separate them, so we proceed with the basis with $\chi_1^\textrm{c}$ and $\chi_2^\textrm{c}$ along with 16, 100, and 200 sweeps of smearing.

\begin{figure}[t]
\includegraphics{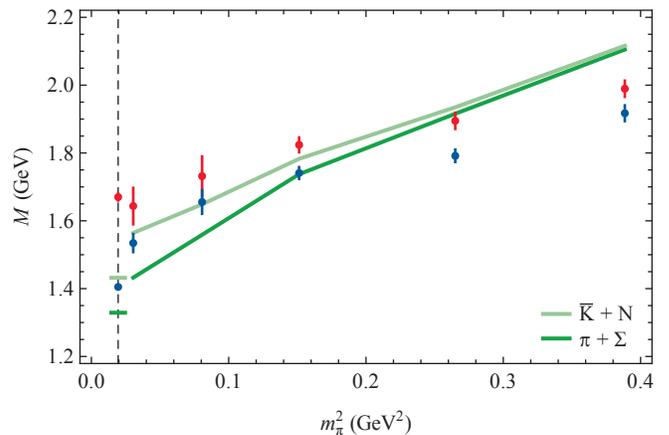}\vspace{-10pt}
\caption{\label{common}(colour online) The lowest-lying eigenstate-projected masses plotted against $m_\pi^2$, along with $\pisigma$ and $\KbarN$ multi-particle scattering thresholds. A correlated error analysis indicates the lowest-lying odd-parity state lies more than one standard deviation below the $\KbarN$ scattering threshold at the lightest quark mass. The ordering of the states is in accordance with Nature.\vspace{-10pt}}
\end{figure}

Now that we have identified the optimal variational parameters and operator basis, we repeat the analysis for the remaining quark masses. By plotting the fits to the lowest-lying projected effective masses against $m_\pi^2$ in Figure~\ref{common}, we can see that our results do trend towards the physical values. However, the lowest-lying state sits too high to approach the $\Lambda$(1405) in the physical limit, especially since finite-size effects are expected to be negligible. As the heavy strange quark interferes with a direct comparison with the physical values, we consider the multi-particle scattering thresholds for the $\pisigma$ and $\KbarN$ states (plotted on the same figure). Our lowest-lying odd-parity state lies between the scattering thresholds as in Nature. This indicates we have isolated the $\Lambda(1405)$.

\begin{figure}[!ht]
\includegraphics{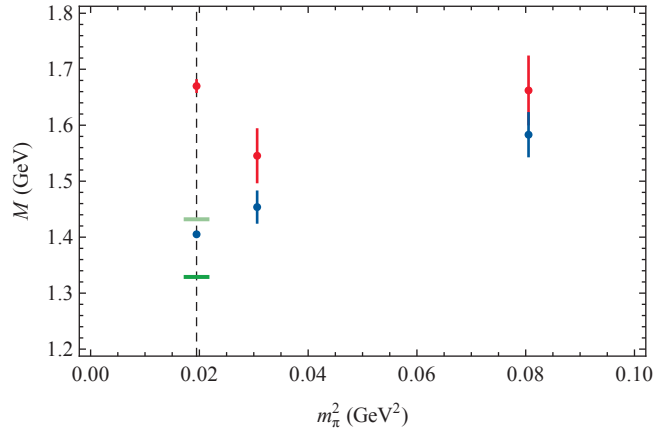}\vspace{-10pt}
\caption{\label{fig:quenched}(colour online) The lowest-lying eigenstate-projected masses after partial quenching of the strange quark to reproduce the physical kaon mass. Extrapolation of the trend for the lowest-lying state to the physical limit reproduces the physical value for the $\Lambda$(1405).\vspace{-20pt}}
\end{figure}

Using the same variational parameters and operator basis as above, we repeat our analysis using the partially quenched strange quark for the lightest two quark masses. This gives us insight into the dependence of each state on the strange quark mass, and allows us to determine if our lowest extracted state approaches the 1405\,MeV in the physical limit. This data is plotted in Figure~\ref{fig:quenched}. As expected, all states have decreased in energy, and now an extrapolation of the trend to the physical limit reproduces the physical value.

\begin{figure}[!ht]
\includegraphics{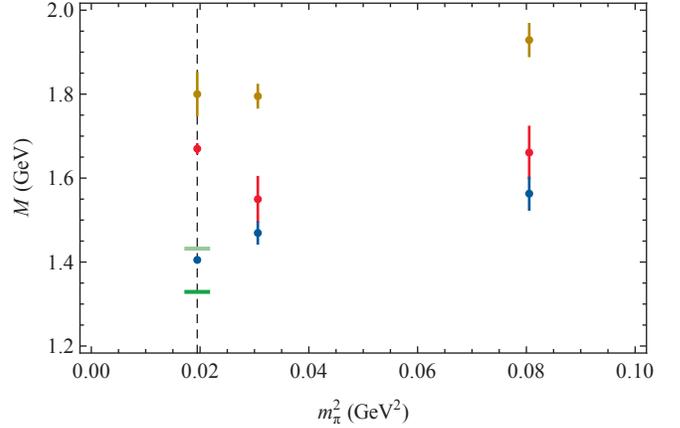}\vspace{-10pt}
\caption{\label{fig:flavour}(colour online) The lowest-lying eigenstate-projected masses (partially quenched) using $\mathop{\textrm{SU}}(3)$-flavour-symmetry-specific operators $\chi^1$, $\chi_1^8$, and $\chi_2^8$ plotted against $m_\pi^2$. Using these operators allows us to isolate three low-lying states.}
\end{figure}

Up to this point, we have used the ``common'' operators, which make no assumptions about the $\mathop{\textrm{SU}}(3)$-flavour-symmetry properties. While we have identified two of the three low-lying states, we are unable to resolve all three. Thus, we extend our analysis to include the conventional, purely octet ($\chi_1^8$ and $\chi_2^8$) and purely singlet ($\chi^1$) operators \cite{Leinweber:1990dv} to isolate all three states. Results are presented from the dimension-9 basis formed from $\chi^1$, $\chi_1^8$, and $\chi_2^8$ at 16, 100, and 200 sweeps of smearing in Figure~\ref{fig:flavour}. Comparing the masses obtained from using just the ``common'' interpolators, we find that the lowest two states remain constant within errors (indicating they are unmixed energy eigenstates), while a third state appears corresponding to the third state in the physical limit. The lowest state is predominately flavour-singlet while the next two states are predominately octet, dominated by $\chi_2^8$ and $\chi_1^8$ for the second and third states, respectively. All results are summarised in Tables~\ref{tab:common} and \ref{tab:flavoured}.

This is the first investigation of the low-lying $J^P = {1/2}^-$ spectrum of the $\Lambda$ baryon at near-physical pion masses, and the first to demonstrate the isolation of the elusive $\Lambda$(1405). Importantly, we have identified a low-lying $\Lambda$(1405) using conventional three-quark operators, and the techniques demonstrated here can be used further to investigate other properties of this unusual state.

\begin{acknowledgments}
This research was undertaken on the NCI National Facility in Canberra, Australia, which is supported by the Australian Commonwealth Government. We also acknowledge eResearch SA for generous grants of supercomputing time. This research is supported by the Australian Research Council.
\end{acknowledgments}

\begin{table}[!ht]
\caption{\label{tab:common}Fits to projected effective masses from $\chi^\textrm{c}$ interpolators before (top) and after (bottom) partial quenching of the strange quark. Values are in GeV.}
\begin{ruledtabular}
\begin{tabular}{cccc}
$m_\pi$ & $\kappa_\textrm{s}$ & $m_1$ & $m_2$ \\
\hline
0.6233(7)  & 0.13640 & 1.917(23) & 1.990(24) \\
0.5148(7)  & 0.13640 & 1.792(18) & 1.895(24) \\
0.3890(10) & 0.13640 & 1.741(17) & 1.824(22) \\
0.2839(12) & 0.13640 & 1.656(35) & 1.732(58) \\
0.1745(22) & 0.13640 & 1.534(26) & 1.643(54) \\[5pt]
0.2839(12) & 0.13694 & 1.583(38) & 1.662(60) \\
0.1745(22) & 0.13694 & 1.454(27) & 1.545(47)
\end{tabular}
\end{ruledtabular}
%
\caption{\label{tab:flavoured}Fits to projected effective masses from flavour-octet and -singlet interpolators after partial quenching of the strange quark to $\smash{\kappa_\textrm{s} = 0.13694}$. Values are in GeV.}
\begin{ruledtabular}
\begin{tabular}{cccc}
$m_\pi$ & $m_1$ & $m_2$ & $m_3$ \\
\hline
0.2839(12) & 1.563(38) & 1.661(61) & 1.929(38) \\
0.1745(22) & 1.469(25) & 1.550(52) & 1.795(27)
\end{tabular}
\end{ruledtabular}
\vspace{-10pt}
\end{table}

\bibliography{paper}

\end{document}